\documentclass[reprint, amsmath, amssymb, aps, pra]{revtex4-2}

\usepackage{graphicx}
\usepackage{dcolumn}
\usepackage{bm}

\usepackage[version=4,arrows=pgf-filled,
textfontname=sffamily,
mathfontname=mathsf]{mhchem}
\usepackage{hyperref}
\begin{document}
\preprint{APS/123-QED}

\title{Theoretical analysis of optical feedback-controlled emission of dual-wavelength lasers}

\author{Mathieu Ladouce}
\email{mathieu.ladouce@vub.be}
\author{Pablo Marin-Palomo}%
\email{pablo.marin-palomo@vub.be}
\author{Martin Virte}%
\email{martin.virte@vub.be}
\affiliation{Brussels Photonics (B-PHOT), Vrije Universiteit Brussel, Pleinlaan 2, 1050 Brussels, Belgium}

\date{\today}

\begin{abstract}
Dual and multi-wavelength lasers, i.e., lasers with the ability to emit at two or more wavelengths in a controlled fashion, represent an exciting new twist in laser physics. Harnessing the mode competition to control them better remains, however, a challenge. In this work, we numerically explore the effect of optical feedback on the emission properties of a dual-wavelength laser using a rate equation model. We focus on switching capability and investigate the impact of key laser and feedback parameters. 
We connect the emergence of simultaneous emission to a lower cross-saturation between modes, and demonstrate that robust switching can be achieved using a short feedback cavity and a sufficiently strong feedback. In particular and in contrast to previous publications, we highlight that the feedback phase difference between the two modes is not a critical parameter. 
Our results are consistent with recent experimental observations, supporting the relevance of feedback-based control techniques for multi-wavelength lasers. 

\end{abstract}

\maketitle

\section{\label{sec:1-Intro}Introduction}
When subject to optical feedback (OF), semiconductor lasers exhibit various behaviors, ranging from stability enhancement and mode hopping to chaos, that have driven interest for various applications such as linewidth reduction \cite{Chraplyvy1986}, chaos-based secure communication \cite{Argyris2005} or random bit generation \cite{Uchida2008}.
Feedback can also be used for controlling modes in several types of semiconductor laser structures.
For example, changing the feedback conditions plays the role of a selection mechanism between transverse modes \cite{Shore1996, Law1997} or polarization modes \cite{Russell1997, Valle1998} in VCSELs, sometimes showing strong suppression ratios.
In quantum dot lasers, an appropriate level of feedback can trigger stable simultaneous emission of the ground and excited states \cite{Naderi2010}.
Furthermore, adjusting the round-trip time in the feedback cavity enables recurrent switching between the single emission of one of these two states \cite{Virte2014, Pawlus2016}.
Similarly, wavelength selection in quantum well lasers has been demonstrated experimentally and theoretically in \cite{Osborne2012, Pawlus2022}.
These studies suggest that the mode-selective gain amplification due to feedback is crucial in enhancing inherent mode competition mechanisms, like spatial hole burning, to enable feedback-induced emission state control.

Dual-Wavelength lasers (DWLs), and more generally multi-wavelength lasers (MWL), i.e., lasers designed to emit at a specific set of wavelengths in a controllable way, are desirable for several applications \cite{Osborne2012}.
On the one hand, a device that emits two wavelengths simultaneously could be used for high-speed signal generation by photomixing \cite{Tani2005}.
On the other hand, combined with optical injection, discrete switching between wavelength channels could be used for all-optical signal wavelength conversion \cite{White2002} or processing toward future photonic networking \cite{Papadimitriou2003}.
Among the various DWL sources,  the device proposed in \cite{Pawlus2022} is a promising candidate owing to its monolithic integration and built-in control scheme.
It consists of a dual-cavity laser coupled to a phase-controlled OF cavity.
Each cavity is designed for a given wavelength so that the separation between both wavelengths can be adjusted at the design stage.
It has a single gain section, allowing competition for carriers between the lasing modes and thus their selectivity through feedback control.
Furthermore, its on-chip monolithic integration offers major advantages such as compactness, cost reduction, and low energy consumption.
The feedback-based mode control technique consists of biasing a phase shifter (PS) within the feedback cavity.
It allows controlling laser emission with periodic switching between two wavelengths separated by ~1.3 THz with suppression ratios greater than 30 dB \cite{Pawlus2022}. 
In \cite{Abdollahi2024}, a configuration involving three wavelengths switching together with a fourth one remaining unaffected was obtained by carefully tailoring the lasing and feedback conditions.
In combination with optical injection, this mode control scheme allowed for agile multiplication of a narrowband frequency comb over the terahertz range \cite{Abdollahi2024}.

Despite these promising results, the underlying mechanism enabling the emission control remains poorly understood from a theoretical standpoint.
Experimental observations and rate equation-based models have revealed that only under specific conditions of the laser and feedback cavity, transitions between the possible emissions could be observed \cite{Pawlus2022}.
This raises fundamental questions about the interplay of laser and feedback parameters on the feedback-based control mechanism, especially considering the variability of the manufacturing process. 

In this study, we address this gap by exploring, both analytically and numerically, the parameter space of a theoretical model relying on a multimode extension of the Lang-Kobayashi rate equations.
By systematically analyzing key parameters, such as the cross-saturation parameter, modal gain difference, feedback strength, feedback delay, and feedback phase difference, we identify the requirements to obtain the possible emission states of the laser, namely single wavelength emission or simultaneous emission at both wavelengths, when changing the feedback round-trip phase.
We also establish figures of merit characterizing the transitions between those possible emission states and the appearance of dynamical regimes.
In Section ~\ref{sec:3-LaserEffects}, feedback parameters are kept constant and set to values presumably allowing for wavelength switching, i.e., transitions between single emission of the two wavelengths.
We identify the requirements and the signature of the modal gain difference and the cross-saturation parameter on the emission control of the laser.
Afterwards, in Section ~\ref{sec:4-OFBEffects}, we investigate the requirements to obtain either simultaneous emission or wavelength switching in terms of feedback parameter values.
We also expose the appearance of dynamical regimes (periodic oscillations or chaos).
Finally, in Section ~\ref{sec:5-Experiment}, we report wavelength switching measured experimentally from a MWL monolithically co-integrated in the InP platform with a feedback cavity.
The influence of the controllable parameters, namely the modal gains and the feedback strength, on the measured transitions is compared to our numerical predictions.

\section{\label{sec:2-Theory}Theoretical framework}
In this section, we describe in detail the feedback-based emission control mechanism for a DWL, along with the theoretical model and associated variables. 

\subsection{\label{sec:2-1-EmissionControl} Feedback-induced emission control}
\begin{figure}[b]
    \includegraphics{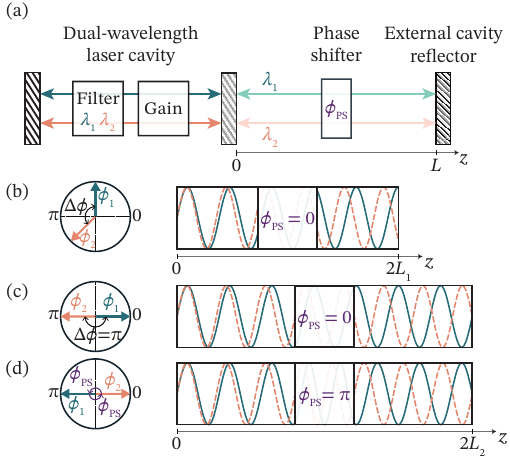}
    \caption{
    (a) Schematic of the feedback-based control setup. The DWL cavity includes a gain section and a filter selecting two wavelengths $\lambda_1$ and $\lambda_2$. The external cavity is formed by a third mirror at a distance $L$ from the laser front facet and includes a phase shifter. 
    (b-d) Phase diagrams (left) show the modal feedback phases accumulated after one round-trip inside the feedback cavity, represented on the right. In (b), a cavity of arbitrary length is considered, while in panels (c,d), the cavity length is tailored such that the two modes are in anti-phase, i.e., $\Delta\phi=\pi$. A phase-shift $\phi_\text{PS}$ of $0$ and $\pi$ is applied in (c) and (d), respectively.
    }
    \label{fig:1-OFBControl}
\end{figure}

The feedback cavity acts as an external Fabry-Pérot cavity through which each mode accumulates a round-trip phase $\phi_m$. In-phase modes will resonate and experience a gain boost, while the modes in anti-phase will experience higher losses \cite{Pawlus2022}. A phase shifter (PS) can then be used to adjust the feedback phase - by adding a phase-shift $\phi_\text{PS}$ - and, thus, to actively select which mode will be resonant as pictured in  Fig.~\ref{fig:1-OFBControl}(a). 

\noindent
For each lasing mode $\lambda_1$ and $\lambda_2$, the total feedback phase $\psi_m$ is equal to $\psi_m = \phi_m + \phi_\text{PS}$ with $\phi_m = 2n_{\text{eff},m} L\frac{2\pi}{\lambda_m}$ where $m$ is the mode number, $n_{\text{eff},m}$ is the effective refractive index of the medium and $L$ is the length of the feedback cavity. Without loss of generality, we subtract $\phi_1$ in both equations, and obtain:
\begin{align}
    \psi_1 &= \phi_{\text{PS}}, \label{eq:psi1} \\
    \psi_2 &=  \Delta\phi + \phi_{\text{PS}} \label{eq:psi2}
\end{align}
where $\Delta\phi=\phi_2-\phi_1$ is the feedback phase difference. At this stage, we assume that the phase shift added by the modulator is identical for all wavelengths. We can expect that, for a DWL emitting at $\lambda_1$ and $\lambda_2$, obtaining wavelength switching will be the most effective when the two wavelengths are in anti-phase after a round trip in the external cavity, i.e., when the phase difference $\Delta\phi = \pi$, see Fig.~\ref{fig:1-OFBControl}(c-d) \cite{Pawlus2022}. By adjusting the cavity length, e.g., from $L_1$ to $L_2$, see Fig.~\ref{fig:1-OFBControl}(b,c), we can adjust the phase difference $\Delta \phi$. In this configuration, setting $\phi_\text{PS} = 0$ would then make $\lambda_1$ resonant and $\lambda_2$ anti-resonant as in (c). Setting $\phi_ \text{PS} = \pi$ reverses the situation and results in a gain boost for $\lambda_2$ over $\lambda_1$. \\
We will first consider this presumably ideal case of $\Delta\phi = \pi$ in Section~\ref{sec:3-LaserEffects}, while the effect of a non-ideal value of the phase difference will be investigated in Section~\ref{sec:4-OFBEffects}.

\subsection{\label{sec:2-2-EmissionFoMs}Charaterizing the DWL emission}
To analyze the DWL emission properties when varying the feedback phase $\phi_\text{PS}$, we propose here several figures of merit (FoM).\\

First, we distinguish two exclusive states: single and simultaneous emission. We consider that the laser is in a single emission state if the power ratio expressed in dB between the two modes exceeds 10 dB: $|10\log_{10}(P1/P2)|\geq 10$ dB.
Below that threshold, we consider the laser to emit simultaneously at the two wavelengths. We define two suppression ratios: 
\begin{equation}
R_{1,2} = \max_{\phi_\text{PS}} \frac{P_1}{P_2}, \; R_{2,1} = \max_{\phi_\text{PS}} \frac{P_2}{P_1}.
\label{eq:SuppressionRatios}    
\end{equation}
Since $R_{1,2}$ and $R_{2,1}$ can take largely different values, we define the switching suppression ratio $R_\text{switch}$ as the smallest of these two values $R_\text{switch} = \text{min}(R_{1,2}, R_{2,1})$. In this work, we talk about a "complete switch" when $R_\text{switch}\geq 10$ dB, i.e., when both modes can be suppressed by at least 10 dB compared to the other.\\

\begin{figure}
    \includegraphics[width=1\linewidth]{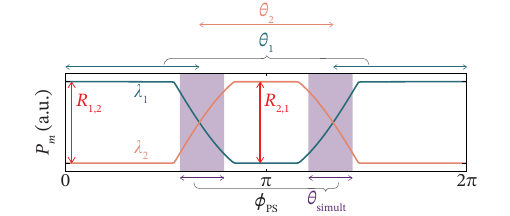}
    \caption{
    Schematic evolution of modal power $P_m$ as a function of $\phi_\text{PS}$.
    $\theta_1$ and $\theta_2$ represent the accumulated range of $\phi_\text{PS}$ values where the first and the second mode are dominant, respectively. $R_{1,2}$ and $R_{2,1}$ are the maximum modal suppression ratios. $\theta_\text{simult}$ gives the cumulative range of $\phi_\text{PS}$ for which simultaneous emission is obtained, shown in purple.
    }
    \label{fig:2-FoMs}
\end{figure}

Second, we define $\theta_m$ as the $\phi_\text{PS}$ cumulative range where the mode at wavelength $\lambda_m$ is dominant within the $2\pi$ span, see Fig.~\ref{fig:2-FoMs} for a graphical representation. The emission asymmetry is given by the difference  $\theta_1 -  \theta_2$. In the extreme cases where the modes at wavelengths $\lambda_1$ or $\lambda_2$ are dominant for all $\phi_\text{PS}$, we have  $\theta_1 -  \theta_2=2\pi$ and $\theta_1 -  \theta_2=-2\pi$, respectively. If both modes are dominant for a similar cumulative range of $\phi_\text{PS}$, then  $\theta_1 - \theta_2=0$. By definition, $\theta_1 + \theta_2  = 2\pi$, as pictured in Fig.~\ref{fig:2-FoMs}. \\
We also define $\theta_\text{simult}$ as the cumulative range for the controlled phase term $\phi_\text{PS}$ in which the state of the laser corresponds to simultaneous emission. Again, the value of $\theta_\text{simult}$ ranges from 0 to $2\pi$, see Fig.~\ref{fig:2-FoMs}. \\
Additionally, depending on the configuration of the laser and feedback parameters, the laser can exhibit dynamical behavior such as periodic oscillation or chaos. We define $\theta_\text{dyn}$ as the range of $\phi_\text{PS}$ where at least one mode shows dynamics. 

\subsection{\label{sec:2-3-Model}Theoretical Model}
To describe multimode semiconductor lasers with optical feedback, we use the model introduced in \cite{Viktorov2000, Koryukin2003}, which is a multimode extension of the Lang-Kobayashi rate equations \cite{Lang1980}.
Each mode is described by a set of two rate equations, one for the complex field amplitude $E_m$ and one for the carrier density $N_m$.
Phenomenological mode-mode coupling terms are introduced in the carrier equations with the self and cross-saturation parameters $\beta_{mm}$ and $\beta_{mn}$, respectively. In this work, we make the approximation that the self-saturation parameters are equal to one $\beta_{11} = \beta_{22} = 1$, and that both cross-saturation parameters have the same value $\beta_{12}=\beta_{21}=\beta$ ($0\geq\beta\geq1$). Modal gain terms $g_m$ are also included to express wavelength-dependency of the gain. Equations for each mode are expressed as follows:
\begin{eqnarray}
    \frac{dE_1}{dt} &=& (1 + \mathrm{i} \alpha)
    \left( g_1N_1 - \frac{1-g_1}{2} \right) E_1 \nonumber \\
    && + \kappa \mathrm{e}^{\mathrm{i}\psi_1} E_1(t-\tau)
    + \xi\sqrt{\beta_\text{sp}}, \label{eq:FieldEqMode1} \\
    \frac{dE_2}{dt} &=& (1 + \mathrm{i} \alpha)
    \left( g_2N_2 - \frac{1-g_2}{2} \right) E_2 \nonumber \\
    && + \kappa \mathrm{e}^{\mathrm{i}\psi_2} E_2(t-\tau)
    + \xi\sqrt{\beta_\text{sp}}, \label{eq:FieldEqMode2} \\
    T\frac{dN_1}{dt} &=& P - N_1 - (1+2N_1) \nonumber \\
    && \times (g_1\beta_{11}|E_1|^2 + g_2\beta_{12}|E_2|^2), \label{eq:CarrierEqMode1} \\
    T\frac{dN_2}{dt} &=& P - N_2 - (1+2N_2) \nonumber \\
    && \times (g_1\beta_{21}|E_1|^2 + g_2\beta_{22}|E_2|^2), \label{eq:CarrierEqMode2}
\end{eqnarray}
with $\alpha$ being the linewidth enhancement factor, $\kappa$ the feedback strength, $\tau$ the feedback delay and $P$ the normalized injection current $(P=(J-J_\text{th})/(2J_\text{th}))$, with $J$ being the injection current and $J_\text{th}$ the threshold current. These equations are normalized with respect to the photon lifetime $\tau_\text{p}$, with $T$ being the carrier to photon lifetime ratio. 
While the feedback phase $\psi_m$ is physically linked to the time delay $\tau$, the delay variation associated with a $2\pi$ phase shift achieved with a phase shifter is negligible compared to the total round-trip time. Therefore, we assume in this work that the two quantities, $\psi_m$ and $\tau$, can be tuned independently.
We apply  equations \ref{eq:psi1} and \ref{eq:psi2} to equations \ref{eq:FieldEqMode1} and \ref{eq:FieldEqMode2} , respectively. 
The field equations also include a spontaneous emission noise term $\xi\sqrt{\beta_\text{sp}}$  with $\beta_\text{sp}$ being the noise factor and $\xi$ a random variable following a complex gaussian distribution.
Unless stated otherwise, we will use the parameter values listed in Table~\ref{tab:FixedParameters}.

\begin{table}[h]
    \caption{Parameters used in equations with their respective domain or values, which are fixed unless explicitly mentioned.}
    \label{tab:FixedParameters}
    \begin{ruledtabular}
    \begin{tabular}{ll}
        Linewidth enhancement factor $\alpha$ & 3\\
        Spontaneous emission noise factor $\beta_\text{sp}$ & $10^{-12}$\\
        Carrier to photon lifetime ration $T$ & 1000\\
        Normalized injection parameter $P$ & 0.5\\
        Self-saturation parameter $\beta_{mm}$ & 1\\
    \end{tabular}
    \end{ruledtabular}
\end{table}

\subsection{\label{sec:2-4-SteadyStates}Steady states solutions for weak feedback}
In this section, we derive analytical solutions for a dual-wavelength laser with a weak feedback $\kappa \ll 1$. In this case, we neglect the spontaneous emission noise term. 
From the two carrier equations (\ref{eq:CarrierEqMode1}, \ref{eq:CarrierEqMode2}), we obtain 4 possible steady-state solutions:
\begin{enumerate}
    \item No emission: $|E_1|^2=0$ and $|E_2|^2=0$
    \item Single emission of mode 1: $|E_1|^2=\frac{P-N_1}{g_1\beta_{11}(1+2N_1)}$ and $|E_2|^2=0$.
    \item Single emission of mode 2: $|E_1|^2=0$ \newline and $|E_2|^2=\frac{P-N_2}{g_2\beta_{22}(1+2N_2)}$.
    \item Simultaneous emission: $|E_1|^2>0$ \newline and $|E_2|^2>0$. 
\end{enumerate}
While the first case will obviously be left aside, for cases 2 and 3, we directly obtain an expression of the field intensity as a function of the carrier population $N_1$ or $N_2$. In the last case, however, we obtain the following equations: 
\begin{equation}
    \begin{pmatrix}
        g_1\beta_{11} & g_2\beta_{12} \\
        g_1\beta_{21} & g_2\beta_{22}
    \end{pmatrix}
    \begin{pmatrix}
        |E_1|^2 \\
        |E_2|^2        
    \end{pmatrix}
    =
    \begin{pmatrix}
        \frac{P-N_1}{1+2N_1} \\
        \frac{P-N_2}{1+2N_2}
    \end{pmatrix}
    \label{eq:SimultaneousEmissionSystem}
\end{equation}
Since both amplitudes need to be real and positive, this set of equations only accepts a solution if the determinant of the 2x2 matrix on the left is strictly positive, i.e., $\beta_{11}\beta_{22}>\beta_{12}\beta_{21}$. In other words, we can only have simultaneous emission if self-saturation is stronger than cross-saturation. Under this condition, the field intensities are given by:
\begin{eqnarray}
    && \begin{pmatrix}
        |E_1|^2 \\
        |E_2|^2        
    \end{pmatrix}
    = \nonumber \\
    && \frac{1}{\beta_{11}\beta_{22}-\beta_{12}\beta_{21}}
    \begin{pmatrix}
        \frac{\beta_{22}(P-N_1)}{g_1(1+2N_1)} - \frac{\beta_{21}(P-N_2)}{g_1(1+2N_2)} \\
        \frac{\beta_{11}(P-N_2)}{g_2(1+2N_2)} - \frac{\beta_{12}(P-N_1)}{g_2(1+2N_1)}
    \end{pmatrix}
    \label{eq:SimultaneousEmissionAmplitudes}
\end{eqnarray}

To find an expression for $N_1$ and $N_2$, we do a standard phase-amplitude decomposition of both field equations: $E_m(t) = A_m(t){e}^{i\Psi_m(t)}$ 
and derive their corresponding equations. For steady states, we have $\frac{dA_m}{dt} = 0$ and $\frac{d\Psi_m}{dt} = \Delta_m$ with $\Delta_m$  the frequency shift of the mode compared to the free-running case. 
We then obtain the two following equations:
\begin{gather}
    0 = \left( g_m N_m - \frac{1-g_m}{2} \right) A_m \nonumber \\
    + \kappa A_m (t-\tau) \cos\left(\Psi_m (t-\tau)
    - \Psi_m(t) + \psi_m \right), \label{eq:RealPartSteadyStatesEquations} \\
    \Delta_m = \alpha \left( g_m N_m - \frac{1-g_m}{2} \right) A_m \nonumber \\
     + \kappa \frac{A_m (t-\tau)}{A_m(t)} \sin\left(\Psi_m (t-\tau)
    - \Psi_m(t) + \psi_m \right). \label{eq:ImPartSteadyStatesEquations}
\end{gather}

Finally, when assuming $\kappa \ll 1$, we can show (derivation details are reported in the appendix) that
\begin{eqnarray}
    \Delta_m &=&
    \frac{-\kappa(\alpha\cos\psi_m - \sin\psi_m)}
    {1+\tau\kappa(\alpha\sin\psi_m + \cos\psi_m)},
    \label{eq:SteadyFrequencyShift}
    \\
    N_m &=& \frac{1-g_m}{2g_m} - \frac{\kappa}{g_m} \cos(-\Delta_m\tau + \psi_m).
    \label{eq:SteadyCarrierPopulations}
\end{eqnarray}
In practice, as described in the next sections, we use these steady-states as a starting point for simulations and compute their stability with DDE-BIFTOOL v.3.1 \cite{Sieber2016}.

\begin{figure}[b]
    \includegraphics[width=1\linewidth]{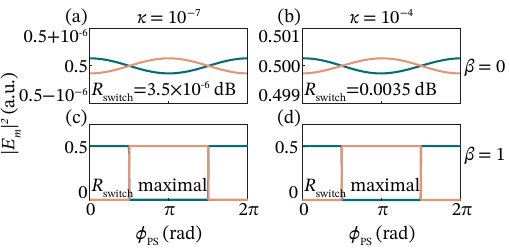}
    \caption{Output power of the two laser wavelengths derived from the steady state solutions for $\kappa = 10^{-7}$ (a, c) and $\kappa = 10^{-4}$ (b, d). Panels (a, b) are with $\beta=0$, panels (c, d) with $\beta=1$.}
    \label{fig:3-NecessaryCoupling}
\end{figure}

\section{\label{sec:3-LaserEffects} Modal gain and cross-saturation effects on the DWL emission}
While the control technique relies on tuning the feedback phase, the mode coupling - modeled by the cross-saturation parameter $\beta$ - is an essential ingredient as well. In Fig.~\ref{fig:3-NecessaryCoupling}, we show that the lack of coupling prevents any significant switching between modes. We use $g_1=g_2$ to maintain symmetry between modes and arbitrarily select $\tau=100$ though other values lead to the same outcome. Interestingly, with a strong coupling $\beta=1$, we observe a switching with maximal amplitude even for very small feedback rates. Additionally, lasers with unbalanced gains $g_1 \neq g_2$ will unavoidably favor the mode with higher gain, potentially to the point where switching or even simultaneous emission might become impossible. \\
In this section, we therefore focus on the laser parameters and investigate the influence of the modal gain difference $\Delta g = g_1 - g_2$ and the cross-coupling parameter $\beta$ on the proposed feedback-based control technique. The feedback parameters are, on the other hand, set to values presumptively suited for wavelength switching, in particular $\Delta \phi = \pi$ and $\kappa=10^{-4}$. 

\begin{figure}[b]
    \includegraphics[width=1\linewidth]{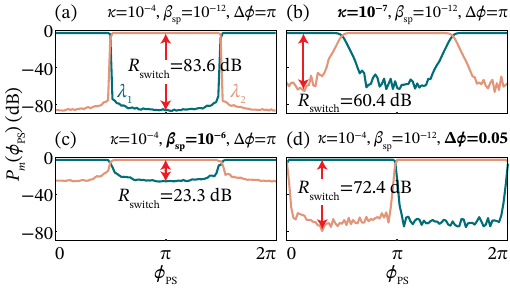}
    \caption{Output powers evolution for a perfect DWL for different noise and feedback parameters. (a) Baseline with default parameters $\kappa = 10^{-4}$, $\beta_\text{sp}$ and $\Delta\phi = \pi$. Only one parameter is varied compared to the baseline in the other panel: $\kappa = 10^{-6}$ (b), $\beta_{sp} = 10^{-7}$ (c), and $\Delta\phi = 0.05$ (d). The delay $\tau$ is arbitrarily fixed at 100.}
    \label{fig:4-PerfectDWLSim}
\end{figure}
\subsection{\label{sec:3-1:PerfectDWL} The perfect DWL \texorpdfstring{$(\Delta g = 0 \text{ and }  \beta=1)$}{Delta g = 0 and beta = 1}}
Intuitively, a DWL with perfectly balanced ($\Delta g = 0$) and perfectly coupled modes ($\beta=1$) would be the ideal candidate to achieve switching. 
Here, we study this particular case, relying first on the analytical solutions at the steady state, and then on numerical simulations. \\
Since we impose $\beta = 1$, we have an indeterminacy appearing in Eq.~\ref{eq:SimultaneousEmissionSystem} forbidding simultaneous emission solutions, thus the laser can only emit at a single wavelength.
The two modes have the same gain $g_1=g_2=1$; consequently, the only difference between the respective amplitudes will arise from the feedback selective amplification induced by the difference between the feedback phases $\psi_1$ and $\psi_2$ since $\Delta\phi \neq 0$. As a result, for $\kappa > 0$, a complete switching is systematically achieved when varying $\phi_\text{PS}$. As shown in Fig.~\ref{fig:3-NecessaryCoupling}(c,d), by increasing $\beta$ from 0 to 1 for both $\kappa = 10^{-7}$ and $\kappa = 10^{-4}$, we observe sharp switchings with a switching suppression ratio $R_\text{switch}$ reaching an infinite value as the suppressed mode has a power of zero. This is, of course, only possible in the absence of noise. \\
For simulations, we include a spontaneous emission noise term, but we obtain the same results as detailed in Fig.~\ref{fig:4-PerfectDWLSim}. With a noise parameter $\beta_\text{sp} = 10^{-12}$ and a feedback rate of $\kappa = 10^{-4}$, we observe again sharp switches when varying $\phi_\text{PS}$ but with a finite suppression ratio $R_\text{switch} = 84 \, \text{dB}$ due to the presence of noise as shown in panel (a). The suppression ratio depends directly on the noise parameter, and a stronger noise level will deteriorate the suppression ratio, see panel (c). On the other hand, we also observe that high suppression ratios $R_\text{switch} > 60 \, \text{dB}$ are still recorded even for very low feedback rates $\kappa = 10^{-7}$ (b) or feedback phase difference as low as $\Delta \phi = 0.05$ (d). When $\Delta \phi = 0$, no emission control through feedback is observed, and the modal amplitudes depend only on the initial conditions.\\
In general, low feedback rates result in weak mode selectivity, favoring simultaneous emission. However, when low feedback rates are combined with $\beta=1$, which prevents simultaneous emission, the laser exhibits long transients, significantly increasing the computation time required to reach a stable steady state. We will therefore limit our investigation to $\kappa>10^{-4}$, which is in line with realistic feedback levels.
\begin{figure}[b]
    \includegraphics[width=1\linewidth]{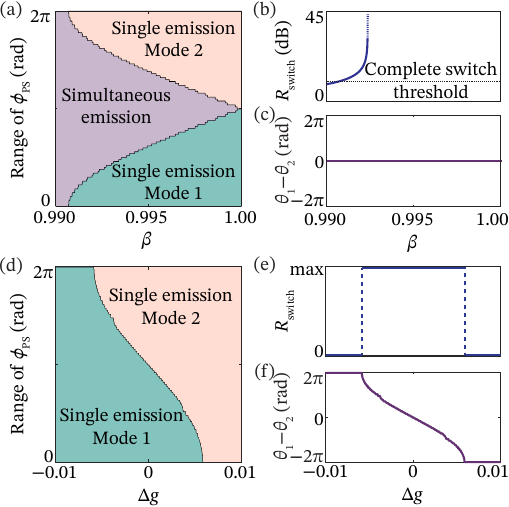}
    \caption{
        Influence of $\beta$ and $\Delta g$ on the DWL emission control.
        (a-c) Influence of the cross-saturation parameter on the FoMs, with $\Delta g=0$.
        (d-f) Influence of the modal gain difference on the FoMs, with $\beta=1$.
        We arbitrarily fixed $\tau=100$ and $\kappa = 10^{-4}$.
        Results were obtained from the steady state solutions.
        }
    \label{fig:5-BetaGainDeviation}
\end{figure}

\subsection{\label{sec:3-2-DeviationFromPerfectDWL} Effects of small deviation from the perfect DWL}

\begin{figure*}
    \includegraphics[width=1\linewidth]{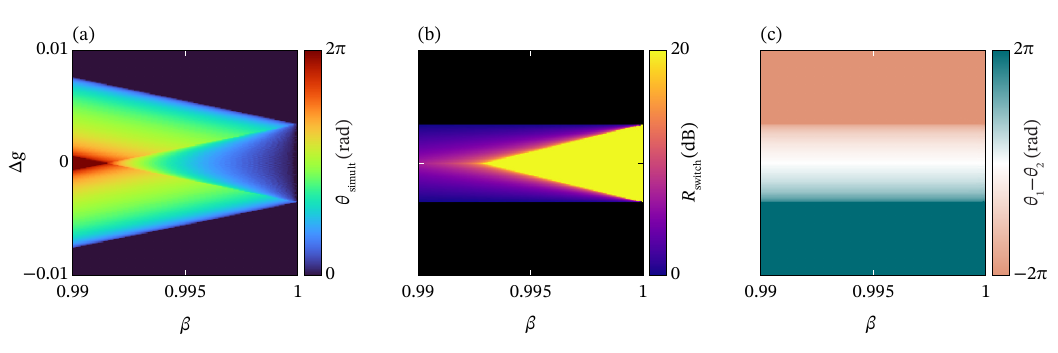}
    \caption{
        Cross-saturation and modal gain effects on the presence of simultaneous emission $\theta_\text{simult}$ (a), the switching suppression ratio $R_\text{switch}$ (b), and the emission asymmetry $\theta_1 -  \theta_2$ (c).
        The delay is set to $\tau=200$ to avoid dynamics, and the strength is fixed to $\kappa = 10^{-3}$ arbitrarily.
        Black areas in the map of $R_\text{switch}$ represent regions where no switching is observed.
        }
    \label{fig:6-BetaGainDeviationMaps}
\end{figure*}

In this section, we deviate from the perfect DWL and explore how non-ideal values of the cross-saturation and gain parameters, i.e., $\beta < 1$ and $\Delta g \neq 0 $, impact the laser behavior when subject to phase-controlled feedback.\\
\textbf{Cross-saturation parameter}: When we deviate from the perfect case by reducing the value of $\beta$ while keeping identical gains $g_1=g_2=1$, we see a progressive appearance of simultaneous emission along the scanned range of $\phi_\text{PS}$ as shown in Fig.~\ref{fig:5-BetaGainDeviation}(a). Below $\beta\approx0.9907$, only simultaneous emission is present, meaning that none of the modes are suppressed by more than 10 dB for any value of $\phi_\text{PS}$, see Fig.~\ref{fig:5-BetaGainDeviation}(b). On the other hand, $R_\text{switch}$ experiences an exponential increase for higher $\beta$ values. We, however, notice that no asymmetry between modes 1 and 2 emerges: $\theta_1 -  \theta_2$ remains constant despite variations of $\beta$, as seen in Fig. ~\ref{fig:5-BetaGainDeviation}(c)\\
\textbf{Modal gain difference}: we fix $\beta=1$ and vary the gain difference $\Delta g$. As discussed above, only single emission solutions are permitted in this configuration. We confirm this feature in Fig.~\ref{fig:5-BetaGainDeviation}(d) and observe a clear asymmetry between emission from modes 1 and 2 depending on the gain difference. While complete switching is systematically obtained for small gain differences, for $|\Delta g|>5.83\times10^{-3}$ no switching is observed and only one mode appears for any value of $\phi_\text{PS}$, see also Fig.~\ref{fig:5-BetaGainDeviation}(e). Finally, plotting $\theta_1 -  \theta_2$ as in Fig.~\ref{fig:5-BetaGainDeviation}(f) unambiguously shows the significant asymmetry between both modes.\\
This first analysis, therefore, unveils two important dependencies: while the cross-saturation parameter has a major impact on the emergence of simultaneous emission, it induces no asymmetry. On the other hand, the gain difference is responsible for the asymmetry.

To go a bit further, we compute maps summarizing the impact of both the modal gain difference and the cross-saturation parameter on the appearance of simultaneous emission and on wavelength switching in Fig.~\ref{fig:6-BetaGainDeviationMaps}. We observe in panel (a) that simultaneous emission mainly occurs for lower cross-saturation values provided that the gain difference remains sufficiently small; indeed, large gain differences obviously prevent simultaneous emission. 
On the other hand, as observed in Fig.~\ref{fig:6-BetaGainDeviationMaps}(b,c), if wavelength switching is the desired scheme of emission control, a smaller gain difference is required; $|\Delta g|\leq0.035$ is needed for any $\beta$ in the considered feedback configuration ($\kappa=10^{-3}$, $\tau = 200$). Wavelength switching with strong suppression ratios requires higher cross-saturation values, i.e., good coupling between the two modes: a minimum cross-saturation of $\beta \approx 0.9915$ is necessary to obtain $R_\text{switch}\geq 10$ dB. This threshold also corresponds to the point where $\theta_\text{simult}$ reaches $2\pi$ when reducing $\beta$ for $\Delta g=0$, see the tip of the reddish triangular shape in Fig.~\ref{fig:6-BetaGainDeviationMaps}(a). Finally, we again confirm that $\beta$ has no impact on the asymmetry of the laser response when varying $\phi_\text{PS}$ as shown in panel(c). This is an interesting feature as it allows us to distinguish between the impact of the gain difference and the mode coupling. 

\begin{figure}
    \includegraphics[width=1\linewidth]{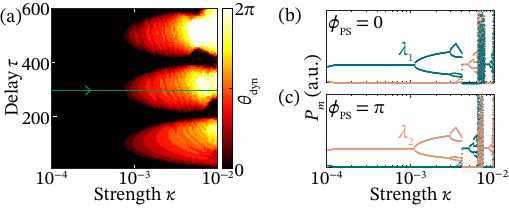}
    \caption{
        (a) Map of $\theta_\text{dyn}$ highlighting the emergence of dynamics as a function of the feedback strength $\kappa$ and delay $\tau$.
        (b,c) Bifurcation diagrams of $P_1$ and $P_2$ as a function of $\kappa$ for $\phi_\text{PS}=0$ (b) and $\phi_\text{PS}=\pi$, with $\tau=300$ and $\Delta\phi$=$\pi$. Laser parameters were set at: $\Delta g=0$ and $\beta=0.999$.
        }
    \label{fig:7-Dynamics}
\end{figure}

\begin{figure*}
    \includegraphics[width=1\linewidth]{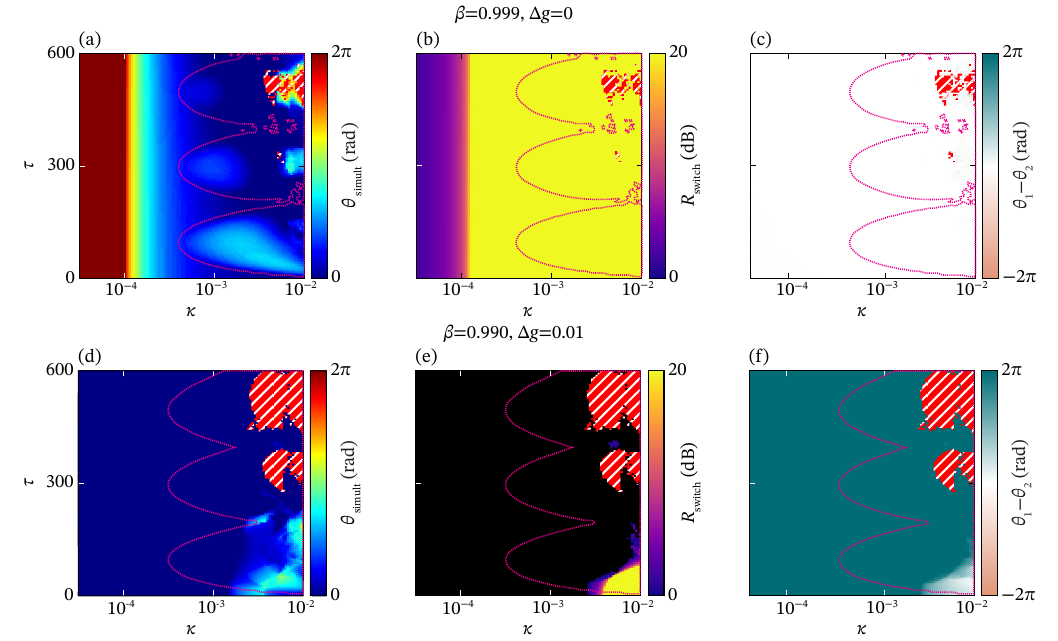}
    \caption{
        Influence of feedback parameters on the simultaneous emission (a, d), the switching ratio (b, e) and the asymmetry (c, f) for two configurations of laser parameters:  an ideal laser (a-c), with $\Delta g=0$ and $\beta=0.999$, and a less ideal laser (d-f), with $\Delta g=0.01$ and and $\beta=0.990$.
        In the closed region delimited by pink dotted lines, we have $\theta_\text{dyn}>0$. Red striped areas correspond to configurations dominated by dynamics, i.e., where $\theta_\text{dyn}=2\pi$. Black areas in the map of $R_\text{switch}$ represent regions where no switching is observed. 
        }
    \label{fig:8-OFBInfluenceFoMMaps}
\end{figure*}

\section{\label{sec:4-OFBEffects}Influence of feedback parameters}
We now investigate the influence of the feedback parameters, namely the feedback rate $\kappa$, delay $\tau$, and the phase difference $\Delta\phi$, on the DWL emission control.
Intuitively, the best control performances will be achieved when the two modes are in an antiphase configuration, i.e., the feedback difference $\Delta\phi = \pi$. Experimental results, however, tend to indicate that such a requirement is not strict. Indeed, convincing performances have been reported despite practical constraints limiting the accuracy of the external cavity design  \cite{Pawlus2022, Abdollahi2024}.
In a first stage, we will therefore analyze the effect of $\kappa$ and $\tau$ in the ideal $\Delta\phi = \pi$ configuration, before investigating the impact of a non-ideal feedback phase difference $\Delta\phi \neq \pi$ on control performances.

\subsection{\label{sec:4-1-DeltaPhiPi} Anti-phase configuration \texorpdfstring{$\Delta\phi = pi$}{Delta phi = pi}}
As expected with optical feedback, complex dynamics can emerge depending on the feedback parameters. To characterize under which feedback conditions dynamics appears, we computed $\theta_\text{dyn}$ as a function of $\kappa$ and $\tau$, see Fig.~\ref{fig:7-Dynamics}, for a given set of laser parameters. In (a), we clearly see that the figure of merit $\theta_\text{dyn}$ increases along with the feedback strength, with a threshold for the onset of dynamics around $\kappa\approx3.16\times10^{-4}$. We also observe well-known periodic stability enhancements with a period of 200 units of photon lifetime for the time-delay $\tau$, matching the relaxation oscillation frequency $f_\text{RO} \approx \frac{1}{2\pi}\sqrt{\frac{2P}{T}}=1/200$. It is, however, important to note that it is only when $\theta_\text{dyn}=2\pi$, shown in white, that the laser exhibits dynamics for all values of $\phi_\text{PS}$. For $0>\theta_\text{dyn}>2\pi$, dynamics occurs for some values of the feedback phase, while others lead to a stable output. Interestingly, from the bifurcation diagrams in Fig.~\ref{fig:7-Dynamics}(b,c) taken for $\phi_\text{PS}=0$ and $\phi_\text{PS}=\pi$, respectively, at $\tau=300$, we observe a re-stabilization of the laser emission and mode switching for feedback rates around $\kappa = 4.1\times 10^{-3}$ and $\kappa = 7.6\times 10^{-3}$. Since $\Delta g = 0$, the two cases are perfectly symmetrical. Naturally, other $\phi_\text{PS}$ values will lead to a different bifurcation diagram with, e.g., dynamics appearing for different values of the feedback strength $\kappa$.\\

\begin{figure*}
    \includegraphics[width=1\linewidth]{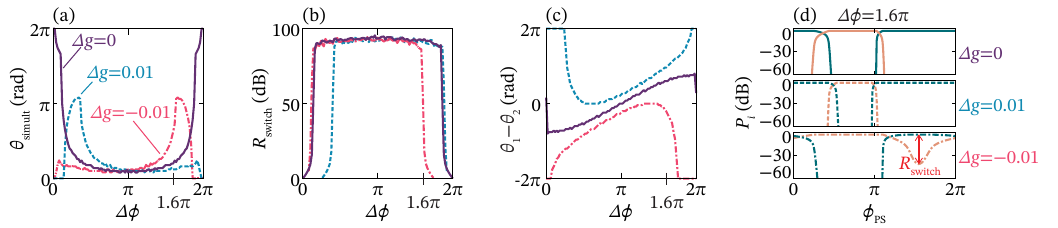}
    \caption{
        Influence of feedback phase difference on simultaneous emission (a), switching ratio (b), and asymmetry (c) for three different laser parameters: $\Delta g=0$ (purple plain lines), $0.01$ (blue dashed line), and $-0.01$ (pink dash-dotted line). Feedback parameters are $\kappa=10^{-3}$, $\tau=200$. Cross-saturation is fixed at $\beta = 0.999$. 
        (d) Examples of modal competition as a function of $\phi_\text{PS}$ when $\Delta\phi=1.6 \pi$ for the same cases. 
        }
    \label{fig:9-DeltaPhiEffect}
\end{figure*}

We consider two lasers with fixed parameters: an ideal DWL, with balanced gains and strong cross-coupling ($\Delta g=0$, $\beta=0.999$), and a less ideal DWL with poor gain balance and cross-coupling ($\Delta g$=0.01, $\beta$=0.990). We calculate the different figures of merit for those two lasers in Fig.~\ref{fig:8-OFBInfluenceFoMMaps}\\
For the ideal DWL ($\beta=0.999$, $\Delta g=0$), as observed in Fig.~\ref{fig:8-OFBInfluenceFoMMaps}(a-c), the laser emission can be fully controlled via tuning of the feedback parameters. Full switching is obtained for all values of $\tau$ provided that $\kappa\gtrsim10^{-4}$. Simultaneous emission can also be obtained, though typically in a lower range of feedback rates $\kappa \leq 3.2\times10^{-4}$ for any delay. For stronger feedback rates, simultaneous emission is driven by delays correlated with the appearance of dynamics. The reader should note that simultaneous emission is also observed in dark blue regions, though it requires fine-tuning of the feedback phase $\phi_\text{PS}$ around the switching point. While dynamics is present for strong feedback and longer delays, it is only in small regions of the feedback parameters that it fully hinders control of the laser emission, see the red area hatched with white stripes. Finally, no asymmetry is observed, which is consistent with the results of the previous section since $\Delta g = 0$. \\
In the case of a less ideal DWL ($\beta=0.990$, $\Delta g=0.01$), we mostly observe single emission without switching in a large range of feedback parameters, see Fig~\ref{fig:8-OFBInfluenceFoMMaps}(d-f). Yet, when combining strong feedback $\kappa \gtrsim 3.5\times 10^{-3}$ and a short cavity $\tau < 100$, simultaneous emission and even complete switching becomes possible. Additionally, dynamics is more present, as shown by the red dashed line indicating the emergence of dynamics for some values of $\phi_\text{PS}$. Furthermore, configurations where dynamics occur for all feedback phase values---thus preventing any control of the laser emission---cover a larger region, though it can be avoided by considering delays below $\tau < 300$.\\
While the preference for short cavities can be easily solved by using co-integrated feedback sections, e.g., in photonic ICs as in \cite{Pawlus2022}, the impact of a small deviation from the ideal case of the laser parameters on the control capability is significant. This aspect is discussed in more detail in section \ref{sec:5-Experiment}, but, in practice, as reported in \cite{Pawlus2022} and \cite{Abdollahi2024}, the feedback control appears to be very effective. This naturally raises the question of what constitutes a realistic parameter range for $\beta$ and $\Delta g$ in real laser structures. 

\begin{figure*}
    \includegraphics[width=1\linewidth]{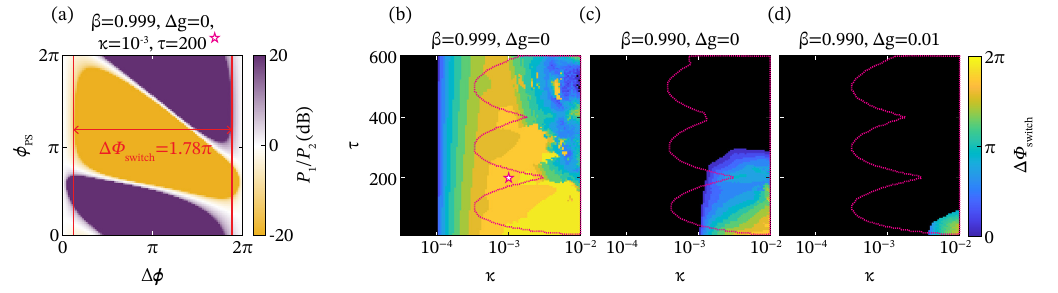}
    \caption{(a) Output power ratio $P_1/P_2$ as a function of $\phi_\text{PS}$ and $\Delta\phi$. Other parameters are: $\beta=0.999$, $\Delta g=0$, $\kappa = 10^{-3}$, $\tau=200$. Vertical red lines correspond to the limits of the range of $\Delta\phi$ values leading to $R_\text{switch}\geq10$ dB. This configuration is shown as a star in panel (b). (b-d) Maps showing $\Delta\Phi_\text{switch}$ as a function of $\kappa$ and $\tau$ for three sets of laser parameters: $\beta=0.999$, $\Delta g =0$ (b), $\beta=0.990$, $\Delta g =0$ (c) and $\beta=0.990$, $\Delta g =0.01$ (d). Black areas represent regions where no switching is observed. The dotted red line delimits the region where dynamics occur. 
        }
    \label{fig:10-DeltaPhiRange}
\end{figure*}

\subsection{\label{sec:4-2-DeltaPhiInfluence}Effects of the phase difference}
Up to now, we assumed that the two modes of the DWL are coming back to the laser cavity in anti-phase after a round-trip in the feedback section, i.e., that $\Delta\phi = \pi$. Yet, this is a strict assumption that is not realistic in most experimental settings, including for integrated systems, due to manufacturing tolerances. Thus, in this section, we study the impact of the feedback phase difference on the DWL emission control mechanism and its figures of merit.\\

As a first approach, we study how the feedback phase difference $\Delta\phi$ impacts the DWL emission in a fixed feedback setup ($\kappa=10^{-3}$ and $\tau=200$) that provides good switching performances when $\Delta\phi = \pi$, see Fig.~\ref{fig:9-DeltaPhiEffect}.\\
We consider a DWL with ideal parameters ($\beta=0.999$ and $\Delta g=0$), and for each value of $\Delta\phi$ between $0$ and $2\pi$, we scan values of $\phi_\text{PS}$ to evaluate our three figures of merit $\theta_{simult}$, $R_\text{switch}$, $\theta_1 -  \theta_2$. For $\Delta\phi = \pi$, switching is observed with almost no simultaneous emission, a strong suppression ratio, and no asymmetry between the modes. When increasing or decreasing $\Delta\phi$, switching performances remain excellent albeit with a moderate increase of simultaneous emission and asymmetry, until a sudden drop in $R_\text{switch}$ is observed for $\Delta\phi > 1.9 \pi$ or $< 0.1\pi$, i.e., when the phase difference gets closer to 0. Understandably, because the two modes $\lambda_1$ and $\lambda_2$ are no longer in anti-phase, the external cavity is no longer providing a sufficiently different forcing to induce a switch. We observe, however, that this only occurs when the feedback phase difference $\Delta\phi$ becomes very small. \\
If we now add slight modal gain differences $\Delta g = \pm 0.01$, we introduce an asymmetry between the modes. We then see in Fig.~\ref{fig:9-DeltaPhiEffect} that the range of good switching performances is reduced for the higher or lower $\Delta\phi$ values. While a larger range of $\phi_\text{PS}$ values leading to simultaneous emission is observed in Fig.~\ref{fig:9-DeltaPhiEffect}(a), along with a stronger asymmetry in Fig.~\ref{fig:9-DeltaPhiEffect}(c), the switching performances remain at the same level, although within a smaller range of $\Delta\phi$ values. For $\Delta\phi \approx 0$, instead of simultaneous emission, single-mode emission from mode 1 or 2, depending on the sign of $\Delta g$, is obtained. The range of $\Delta\phi$ values allowing for complete wavelength switching is still surprisingly large in contrast to the initial assumption that suggests a strict anti-phase configuration of the modes, i.e. $\Delta\phi=\pi$.\\
For the studied feedback configuration, we therefore see that obtaining wavelength switching is possible in a large range of $\Delta\phi$ values around $\pi$. In other words, the proposed control scheme appears to be quite robust against variations of the phase difference. 

Next, we describe how the range of $\Delta \phi$ values leading to switching varies when varying the feedback parameters, and how much one can deviate from the ideal configuration $\Delta \phi = \pi$ and retain a good control over the laser emission.\\ 
In Fig.~\ref{fig:10-DeltaPhiRange}(a), we display the ratio $P_1/P_2$ of the output power of the two wavelengths of the laser in the $(\Delta\phi, \phi_\text{PS})$ plane for an ideal laser configuration with $\kappa = 10^{-3}$ and $\tau=200$. For a given laser and feedback configuration, one can essentially move along the vertical axis by tuning the feedback phase $\phi_\text{PS}$. We can thus observe that good switching is obtained for a large range of $\Delta\phi$. The two red vertical lines indicate the limit cases beyond which $R_\text{switch}<10 \, \text{dB}$, i.e., it is no longer possible to suppress either mode by 10 dB or more. We can then define $\Delta\Phi_\text{switch}$ as the range of $\Delta\phi$ values leading to switching: here $\Delta\Phi_\text{switch} = 1.78 \pi$. In practice, $\Delta\Phi_\text{switch}$ can also be seen as a tolerance measure of how much we can deviate from $\Delta\phi = \pi$. \\
For a broader overview of the effect of feedback parameters, we compute $\Delta\Phi_\text{switch}$ in the $(\kappa, \tau)$ plane for three different lasers, see Fig.~\ref{fig:10-DeltaPhiRange}(b-d). For an ideal DWL, see Fig.~\ref{fig:10-DeltaPhiRange}(b), we confirm again that a minimum feedback strength of $\kappa \approx 10^{-4}$ is necessary to obtain complete wavelength switching even when $\Delta\phi \neq \pi$. As we increase $\kappa$, $\Delta\Phi_\text{switch}$, that is, the range of $\Delta\phi$ values allowing for switching increases, reaching almost $2\pi$ depending on the delay. At higher feedback strengths and delays ($\kappa \gtrsim 0.001$, $\tau\gtrsim 100$), more $\Delta\phi$ values lead to dynamics which reduces the value of $\Delta\Phi_\text{switch}$. Once again, we observe that the best control performances, i.e., with the largest values of $\Delta\Phi_\text{switch}$, are obtained for short cavities and strong feedback. 
When considering less ideal laser parameters, see Fig.~\ref{fig:10-DeltaPhiRange}(c,d), we can still observe complete switches with a good tolerance on $\Delta\phi$ but with a growing constraint on the feedback strength and delay: $\kappa \gtrsim 0.001$ \& $\tau \lesssim 300$ for a laser with bad coupling ($\beta=0.990$, $\Delta g=0$) and $\kappa \gtrsim 0.003$ \& $\tau\lesssim 100$ when considering gain difference ($\beta=0.990$, $\Delta g=0.01$). The feedback strength and delay, therefore, have more impact than the feedback phase difference $\Delta \phi$. \\
Overall, what seemed initially as a key requirement to have the two modes in anti-phase after a round trip in the external cavity, i.e. $\Delta\phi = \pi$, appears to be, in fact, only a good optimum with a relatively large tolerance. This is, naturally, encouraging considering this parameter can be difficult to adjust experimentally. In practice, this result is also consistent with experimental reports where good control performances were obtained in various conditions and for additional wavelengths \cite{Abdollahi2024}. 

\begin{figure*}
    \includegraphics[width=1\linewidth]{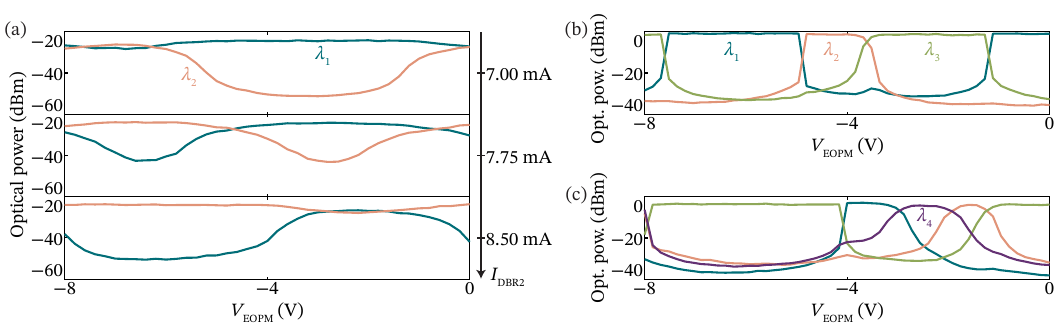}
    \caption{Example of experimental measurements. (a) Output power of the two wavelengths emitted by the laser at $\lambda_1=1537.8$ nm and $\lambda_2=1548.7$ nm as a function of the EOPM voltage $V_\text{EOPM}$ for a DBR current of 7, 7.75, and 8.5 mA from top to bottom. 
        (b-c) Output power of three and four wavelength emitted by the laser at $\lambda_1=1536.7$ nm, $\lambda_2=1536.9$ nm, $\lambda_3=1547.6$ nm, and $\lambda_4=1547.9$ nm when varying the EOPM voltage. Panels (b) and (c) correspond to the same laser under different injection and DBR current configurations.}
    \label{fig:11-Experiment}
\end{figure*}

\section{\label{sec:5-Experiment}Connection with experimental results}
The systematic study performed numerically here can hardly be considered experimentally, as many of the parameters considered are not directly accessible or tunable. Yet, some experimental results can provide an interesting complementary insight. In this section, we report a few experimental measurements using a dual-cavity laser that has shown wavelength switching \cite{Pawlus2022}. Two distributed Bragg reflectors (DBRs) are used for wavelength selection, and they are set at: $\lambda_{1}\approx1548$ nm and $\lambda_{2}\approx1538$ nm. A feedback cavity is co-integrated with the laser and includes a semiconductor optical amplifier (SOA) and an electro-optic phase modulator (EOPM) to control the feedback strength and phase, respectively.\\
Both DBRs can be slightly tuned through current injection, which, in turn, changes the losses of the corresponding mode. As a result, it gives us a way to modify the gain balance between the two modes, similar to a change of $\Delta g$ in our numerical work. We show in Fig.~\ref{fig:11-Experiment}(a) the effect of the DBR tuning on the switching behavior between $\lambda_1$ and $\lambda_2$: we observe a clear transition between a configuration where $\lambda_1$ is dominant to a case where $\lambda_2$ becomes dominant. In between, we obtain a balanced configuration where both modes can be suppressed when varying the feedback phase. Such a result is well in line with the behavior reported in Fig.~\ref{fig:5-BetaGainDeviation}.\\
In a different laser, we were able to observe emission from multiple longitudinal modes in both cavities as shown in  Fig.~\ref{fig:11-Experiment}(b) and (c). Different DBR currents have been used in these two configurations, which makes these two results hard to compare. However, we still observe a clear selective switching between the 3 and 4 modes, in panels (b) and (c) respectively, with a good suppression ratio of other wavelengths of at least 15 dB and typically above 20 dB. Such a result is therefore a clear indication that $\Delta\phi = \pi$ is not a strict requirement and that much smaller feedback phase differences are exploitable in real systems.\\
Unfortunately, the behavior reported here is only observed in a small range of parameters, which prevents us from performing a systematic investigation that could be better compared with the previous numerical study. These few experimental measurements are therefore only presented as supporting evidence of the conclusion of our numerical study. 

\section{\label{sec:6-Conclusion}Conclusion}
We have theoretically investigated the impact of a phase-controlled optical feedback on the emission of a dual-wavelength laser. We base our study on a multimode rate equation model including cross-saturation.\\
We successfully derived steady-state solutions for this model in the case of a weak feedback, and showed that no simultaneous emission at two wavelengths is possible when there is a perfect coupling between the two modes ($\beta = 1$). As such, optical feedback can only trigger complete switching between the two wavelengths. 
While a reduced cross-saturation parameter leads to the emergence of simultaneous emission, we also show that asymmetry is primarily linked with a modal gain difference ($\Delta g \neq 0$).\\
Looking at the effect of feedback parameters, we show that short feedback cavities combined with strong feedback ensure the best control capability over the laser emission, even in the case of non-ideal laser parameters. More importantly, we demonstrate that the feedback phase difference is not so critical: while it was initially thought that a feedback phase difference of $\Delta\phi=\pi$ would be a key requirement, our analysis highlights a good robustness of the control performances against this parameter. In fact, achieving strong feedback in a short cavity appears to be more important than precisely targeting a feedback phase difference of $\pi$. This result is of particular importance to achieve control over several wavelengths where the feedback phase difference would obviously need to be reduced to accommodate additional modes. \\
Finally, although a systematic experimental investigation remains out of reach, we presented a few experimental measurements showing a good agreement with the numerical results and supporting the conclusion of this work. In particular, the capability to control multiple wavelengths appears to be within reach, as shown with the two cases showing selective switching among 3 and 4 modes, and is consistent with the robustness against feedback phase difference.\\

Although the model becomes increasingly more complex with additional modes, with an exponential increase in the number of cross-saturation parameters, further numerical investigations remain necessary to better understand the mode competition and specific impacts of optical feedback. Similarly, experimental efforts including additional wavelengths are needed. In this regard, the use of photonic integrated circuits and a generic foundry platform provides a robust and precise solution, albeit at the cost of flexibility and manufacturing tolerances. The free-space alternative, however, represents a real challenge in terms of alignment and robustness. 

\begin{acknowledgments}
This work was supported by the European Research Council (ERC, Starting Grant COLOR’UP 948129, MV), the Research Foundation Flanders (FWO, grants 1530318N, G0G0319N, MV, and postdoctoral fellowship grant 1275924N, PMP), the METHUSALEM program of the Flemish Government (Vlaamse Overheid).
\end{acknowledgments}
\appendix
\renewcommand{\theequation}{A\arabic{equation}}
\setcounter{equation}{0} 
\section*{Appendix: Derivation of steady state solutions for weak feedback}
We detail here the derivation leading to the final expressions of the frequency shift $\Delta_m$ and the carrier density $N_m$ stated in Eq.\ref{eq:SteadyFrequencyShift} and \ref{eq:SteadyCarrierPopulations}, respectively.
At the steady state, we have $\frac{dA_m}{dt}=0$ and $\frac{d\Psi_m}{dt} = \Delta_m$, therefore, it appears that $A_m(t-\tau)=A_m(t)$ and that $\Psi_m(t)=\Delta_mt$ with a constant that we can neglect without loss of generality. The two equations \ref{eq:RealPartSteadyStatesEquations} and \ref{eq:ImPartSteadyStatesEquations} from the main text then become:
\begin{gather}
    0 = g_m N_m - \frac{1-g_m}{2} \nonumber \\
    +  \kappa \cos\left(-\Delta_m\tau+\psi_m\right),
    \label{eq:RealPartSteadyStatesEquationsSimplified} \\
    \Delta_m = \alpha \left( g_m N_m - \frac{1-g_m}{2} \right) \nonumber \\
     + \kappa \sin\left(-\Delta_m\tau+\psi_m\right). \label{eq:ImPartSteadyStatesEquationsSimplified}
\end{gather}
We isolate $\left( g_m N_m - \frac{1-g_m}{2} \right)$ in Eq.\ref{eq:RealPartSteadyStatesEquationsSimplified} and substitute it in Eq.\ref{eq:ImPartSteadyStatesEquationsSimplified} to obtain:
\begin{equation}
    \Delta_m  = -\alpha\kappa\cos(-\Delta_m\tau+\psi_m) 
    + \kappa\sin(-\Delta_m\tau+\psi_m).
    \label{eq:DeltaEq1}
\end{equation}
Using the trigonometric addition formulas
\begin{eqnarray*}
    \cos(a+b) && = \cos(a)\cos(b) - \sin(a)\sin(b),\\
    \sin(a+b) && = \sin(a)\cos(b) + \cos(a)\sin(b), 
\end{eqnarray*}
we get the following equation:
\begin{eqnarray}
   \Delta_m = && \kappa \Big\{ -\alpha \big[ \cos(- \Delta_m\tau)\cos(\psi_m) \nonumber \\
    &&- \sin(-\Delta_m\tau)\sin(\psi_m) \big] \nonumber \\
    && + \sin(-\Delta_m\tau)\cos(\psi_m) \nonumber \\
    &&+ \cos(\Delta_m\tau)\sin(\psi_m) \Big\} \\ 
    \label{eq:DeltaEq2}
    = && - \kappa \Big\{ \alpha \cos( \Delta_m\tau)\cos(\psi_m) \nonumber \\
    && + \alpha \sin(\Delta_m\tau)\sin(\psi_m) \nonumber \\
    && + \sin(\Delta_m\tau)\cos(\psi_m) \nonumber \\
    && - \cos(\Delta_m\tau)\sin(\psi_m) \Big\}. 
    \label{eq:DeltaEq3}
\end{eqnarray}
By factoring out $\cos(\Delta_m\tau)$ and $\sin(\Delta_m\tau)$, the expression becomes:
\begin{eqnarray}
    \Delta_m = && -\kappa\big\{ \cos(\Delta_m\tau)\big(\alpha\cos(\psi_m)-\sin(\psi_m)\big) \nonumber \\
    && + \sin(\Delta_m\tau)\big(\alpha\sin(\psi_m)-\cos(\psi_m)\big)\big\}.
    \label{eq:DeltaEq4}
\end{eqnarray}
With the approximation that we are using a weak feedback $\kappa \ll1$, and considering that the left-hand side is bounded by $\pm\kappa\alpha$, we expect a small frequency shift $\Delta_m\ll1$.
Although the time delay $\tau$ could typically be in the order of 100 ps to 1000 ps, the approximation that $\Delta_m \ll 1$ is very convenient and will be used as a first approximation.
Thus, $\cos(\Delta_m\tau) \approx 1$ and $\sin(\Delta_m\tau) \approx \Delta_m\tau$.
The Eq.\ref{eq:DeltaEq4} then becomes:
\begin{eqnarray}
    \Delta_m = && \kappa \big( \alpha \cos(\psi_m)-\sin(\psi_m)\big) \nonumber \\
    &&- \kappa\Delta_m\tau \big( \alpha \sin(\psi_m)+\cos(\psi_m)\big),
    \label{eq:DeltaEq5}
\end{eqnarray}
which then gives:
\begin{equation}
    \Delta_m = \frac{-\kappa \big( \alpha\cos(\psi_m)-\sin(\psi_m)\big)}{1+\kappa\tau\big( \alpha\sin(\psi_m)+\cos(\psi_m)\big)}.
    \label{eq:DeltaFinalEq}
\end{equation}
Once the frequency shift $\Delta_m$ has been determined, we can use the amplitude equation, see Eq.\ref{eq:RealPartSteadyStatesEquationsSimplified}, to get the expression of the carrier density $N_m$:
\begin{equation}
    N_m = \frac{1-g_m}{2g_m}-\frac{\kappa}{g_m}\cos(-\Delta_m\tau+\psi_m).
    \label{eq:CarrierDensityFinalEq}
\end{equation}

\bibliography{myReferences}

\end{document}